\documentclass{PoS}

\title{H.E.S.S. discovery of very-high-energy gamma-ray emission of
  PKS~1440-389} 

\ShortTitle{H.E.S.S. discovery of PKS 1440-389}

\author{\speaker{H. Prokoph}$^a$, Y. Becherini$^a$, M. B\"ottcher$^b$,
  C. Boisson$^c$, J.-P. Lenain$^d$, and I. Sushch$^b$ 
  for the H.E.S.S.~Collaboration \\ 
  \newline \llap{$^a$}Department of Physics and Electrical Engineering, Linnaeus
  University, 351 95 V\"axj\"o, Sweden \\ 
  \llap{$^b$}Centre for Space Research, North-West University, Potchefstroom
  2520, South Africa \\
  \llap{$^c$}LUTH, Observatoire de Paris, CNRS, Universit\'e Paris Diderot, 5
  Place Jules Janssen, 92190 Meudon, France \\ 
  \llap{$^d$}LPNHE, Universit\'e Pierre et Marie Curie Paris 6,
  Universit\'e Denis Diderot Paris 7, CNRS/IN2P3, 4 Place Jussieu,
  F-75252, Paris Cedex 5, France \\ 
  
  E-mail: \email{heike.prokoph@lnu.se}  \\
 
}

\abstract{Blazars are the most abundant class of known extragalactic
  very-high-energy (VHE, E$>$100 GeV) gamma-ray sources. However, one of
  the biggest difficulties in investigating their VHE emission resides
  in their limited number, since less than 60 of them are known by
  now. 

In this contribution we report on H.E.S.S. observations of the BL
Lac object PKS~1440-389. This source has been selected as target for
H.E.S.S. based on its high-energy gamma-ray properties measured by
{\it Fermi}-LAT. The extrapolation of this bright, hard-spectrum
gamma-ray blazar into the VHE regime made a detection on a relatively
short time scale very likely, despite its uncertain
redshift. H.E.S.S. observations were carried out with the 4-telescope 
array from February to May 2012 and resulted in a clear detection of
the source. Contemporaneous multi-wavelength data are used to
construct the spectral energy distribution of PKS~1440-389 which can
be described by a simple one-zone synchrotron-self Compton model. }

\FullConference{The 34th International Cosmic Ray Conference,\\
		30 July- 6 August, 2015\\
		The Hague, The Netherlands}

\newcommand{\pks}{PKS~1440-389}
\newcommand{\hess}{H.E.S.S.}
\newcommand{\fermi}{{\it Fermi}}
\newcommand{\swift}{{\it Swift}}

\newcommand{\reffig}[1]{Figure~\ref{#1}}  
\newcommand{\reftab}[1]{Table~\ref{#1}}   

\begin{document}

\section{Introduction}

Active galactic nuclei (AGN), in particular blazars, are the most
abundant extragalactic objects detected at very high energies (VHE;
$E>100$~GeV), constituting a third of the known VHE gamma-ray
sources. Due to their fast flux decrease with increasing energy,
observations at VHE are mostly performed with ground-based imaging
atmospheric Cherenkov telescopes (IACTs). 
The small field of view of current IACTs makes new source discoveries
from a large-scale sky survey difficult and therefore the search for 
VHE-emitting blazars has historically involved targeted observations
of source candidates selected based on their radio and X-ray spectra 
\cite{costamante}. With the launch of \fermi\ in June 2008, 
this has partly shifted towards a selection based on extrapolations
of high-energy (HE; 100~MeV$< E <100$~GeV) gamma-ray spectra into
the VHE regime \cite{2010MNRAS.401.1570T}. This lead to several new
blazar discoveries and motivated the choice for targeted observations
of \pks\ with \hess\ and, eventually, its VHE discovery in April 2012
\cite{pksatel}. 

\pks\ (R.A. = $14^h43^m57^s$, decl. = -$39^{\circ}08'39''$, J2000
\cite{jackson2002}) is a bright blazar in the \fermi-LAT energy
regime and has a hard, well-constrained HE spectrum (spectral index:
$\Gamma_{\mathrm{2FGL}} = 1.77\pm0.06$ \cite{2fgl}). 
Its distance was estimated as $z=0.065$ (6dF Galaxy Survey 
\cite{jones2004}), but this redshift value is not included anymore in
the final version of the 6dF catalog due to poor spectral quality
\cite{jones2009}. Despite many measurements in different wavelength
ranges, the redshift of \pks\ remains unknown with the currently best
constraint to be $0.14 < z < 2.2$ \cite{shaw2013}.

\section{\hess\ data analysis and results}

\hess\ is an array of five imaging atmospheric Cherenkov telescopes
located in the Khomas Highland in Namibia which is sensitive to
gamma-ray energies from a few tens of GeV to about 100~TeV 
\cite{highlight}. Observations presented here were taken before the
installation of the fifth telescope for which the array
characteristics are given in \cite{crab2006}. 

\begin{figure}[ht]
  \centering
  \includegraphics[width=0.75\textwidth]{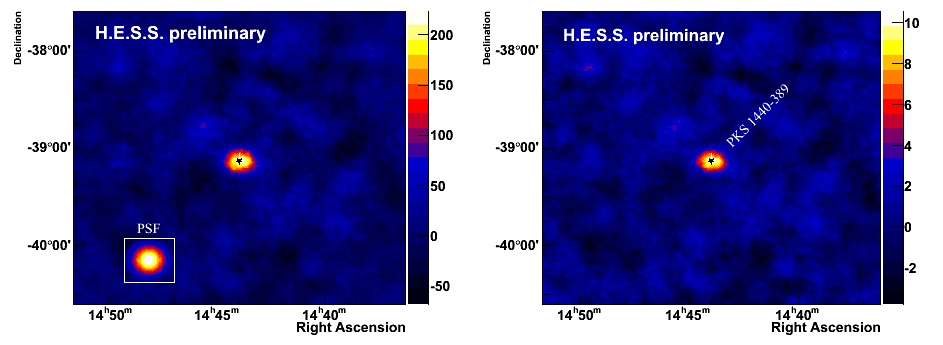}
  \caption{\hess\ sky maps around the position of \pks\ (marked with a star). 
{\it (Left)} The excess map is smoothed with a Gaussian with a width
corresponding to the 68\% confinement radius of the point spread
function (PSF) for this analysis (shown inline). 
{\it (Right)} Corresponding \hess\ significance map. 
} 
  \label{fig:map}
\end{figure}

Observations of \pks\ were taken between 29 Feb and 27 May 2012 at a
mean zenith angle of 17~degrees. All data were taken in wobble mode,
for which the source is observed with an offset of $0.5^{\circ}$ with
respect to the center of the instrument's field of view to allow for
simultaneous background measurements \cite{fomin1994}. After quality
selection and dead time correction, the data sum up to a total live
time of about 12~hours. 
This data set was analyzed with pre-defined cuts, optimized for a
steep-spectrum source as decribed in \cite{parismva}\footnote{Results
  were cross-checked with an independent analysis package \cite{MdN},
  yielding compatible results.}. 
The signal was extracted using a reflected region background model
with an ON region of $0.11^{\circ}$ radius centered on the position of \pks. 
In this ON region, 183 excess events have been detected, corresponding
to a significance of $9.1\sigma$ using Eq.~17 in \cite{lima1983}. 
The excess is found to be consistent with a point-like source 
and the corresponding sky map is shown in Figure~\ref{fig:map}. A fit
to the uncorrelated excess map yields a position for the excess
consistent with the radio position of \pks\ \cite{jackson2002}. 

\begin{figure}[ht]
  \centering
  \includegraphics[width=0.65\textwidth]{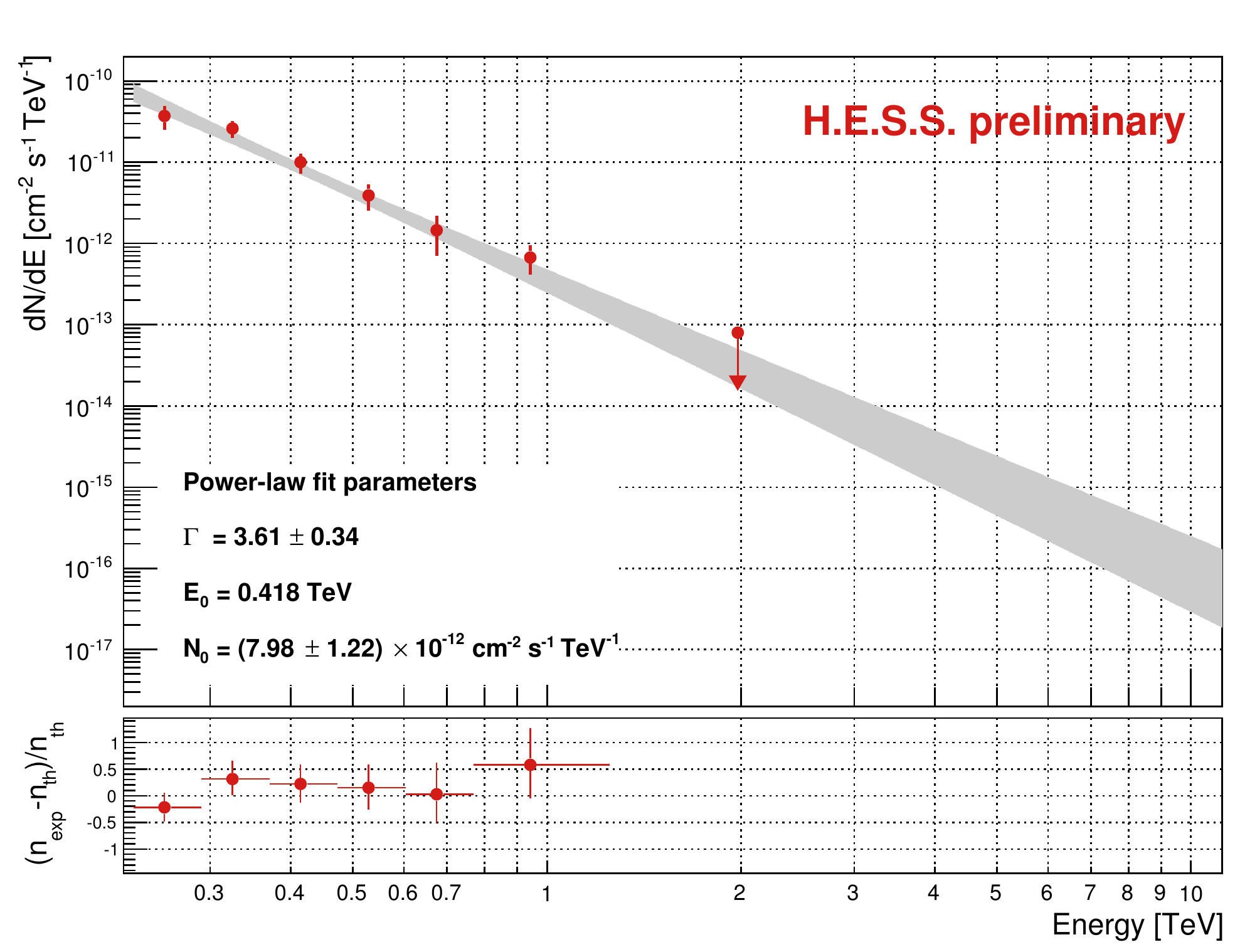}
  \caption{Differential energy spectrum of the VHE gamma-ray emission
    of \pks. Overlaid spectral points were rebinned, requiring a
    minimum point significance of two sigma per bin. 
  }
  \label{fig:spectrum}
\end{figure}

The differential energy spectrum of the gamma-ray emission was 
derived using a forward-folding maximum likelihood fit \cite{piron2001}. 
The photon spectrum, shown in \reffig{fig:spectrum}, is well-described
by a power-law function\footnote{$dN/dE = N_{0} \cdot (E/E_0)^{-\Gamma}$} 
with index $\Gamma = 3.61\pm0.34$ and a flux normalization of $N_{0} =
(7.98\pm1.22) \times 10^{-12} \textrm{cm}^{-2} \textrm{s}^{-1}
\textrm{TeV}^{-1}$ at the decorrelation energy of $E_0 = 418$~GeV
(statistical errors only). 
The integral flux above 220~GeV is $(6.81\pm1.04) \times 10^{-12}
\textrm{cm}^{-2} \textrm{s}^{-1}$, corresponding to about 3\% of the
Crab Nebula flux \cite{crab2006} above the same energy threshold. 
A daily binned light curve was derived for $E>220$~GeV and is shown in
\reffig{fig:lc}. A constant fit to these flux points showed no
significant deviation from a steady flux ($\chi^{2}/ndf = 19.41/14$). 

\section{Multiwavelength data analysis and results}

\subsection{Fermi/LAT}
The Large Area Telescope (LAT) on board the \fermi\ satellite is a
pair-conversion gamma-ray detector, sensitive in the energy range from 
20~MeV to $>300$~GeV \cite{atwood2009}. Data from Aug 2008 to Feb
2015 were analyzed using the LAT ScienceTools (version {\tt v9r33p0})
together with the {\tt P7REP\_SOURCE\_V15} instrument response
functions.  An unbinned likelihood analysis was applied to photon
events with energies from 300~MeV to 300~GeV which were selected in a
circular region of $10^{\circ}$ radius centered on the position of
\pks, resulting in an excess of about $49\sigma$ significance. Two
spectral models were applied: a power law and a log-parabola. 
The results are summarized in \reftab{tab:lat}. 
Using a likelihood ratio test, the log-parabola model was preferred
with respect to the power-law model at the $4.1\sigma$ level. 
A three-monthly binned light curve for the full time range was
computed, resulting in a probability of about 3.4\% for a constant fit 
($\chi^2/ndf = 39.33/25$). 
The calculation of the excess variance \cite{Vaughan2003} showed only a small hint for variability (F$_{var} = 0.13 \pm 0.06$) in the HE regime. 
However, no evidence for variability on a monthly time scale has been
seen within the data set contemporaneous with the \hess\ observations,
shown in \reffig{fig:lc}. 

\begin{table}[htb]
  \centering
  \caption{Results of the spectral fit to the \fermi-LAT data.}
  \label{tab:lat}
  \begin{tabular}{c c c cc}
    \hline \hline
Model & $\alpha$ & $\beta$ & TS & Flux (0.3-300 GeV) \\
&&&&[$10^{-9}$\,ph\,cm$^{-2}$s$^{-1}$]\\
\hline
Log-parabolic (LP) &	$1.32 \pm 0.04$ & $0.083 \pm 0.008$ & 2403.9 &
$8.42 \pm 0.42$\\ 
Power law (PL) & $1.79 \pm 0.03$ & --- & 2411.8 & 
$9.95 \pm 0.53$\\ 
\hline
  \end{tabular}
\end{table}

\subsection{Swift/XRT and UVOT}

The X-ray telescope (XRT) on board the \swift\ satellite is designed
to measure X-rays in the $0.2-10$~keV energy range \cite{burrows2005}. 
Target of opportunity observations were obtained on 2012 April~29 
(MJD~56046), following the VHE discovery of \pks. The analysis of this
8\,ks exposure was performed using standard tools (HEASoft 6.16, Xspec
12.8.2). Data were grouped, requiring a minimum of 20 counts per bin,
and then fitted with a power-law model. 
Photo-electric absorption was fit using the Wisconsin cross-section
\cite{wabs} with a fixed value for the Galactic column density of
$N_{\mathrm{tot}} = 1.08 \times 10^{21}$\,cm$^{-2}$ \cite{NHtot}, 
resulting in a photon index of $\Gamma = 2.64\pm0.05$ and $N_0 =
(3.80\pm0.12)\times10^{-3}$\,cm$^{-2}$\,s$^{-1}$\,keV$^{-1}$ at 1~keV. It
should be noted, that when $N_{\mathrm{tot}}$ was left free during the
fit, a slightly better fit could be obtained with a column density
compatible with the total value. 
 
Simultaneously with the XRT observations, the \swift\ Ultra Violet and
Optical Telescope (UVOT \cite{roming2005}) carried out observations in
all six filters. 
Sky-corrected images were taken from the \swift\ archive, and analysis
was performed using the {\tt uvotmaghist}, {\tt uvotimsum} and {\tt
  uvotsource} tasks included in the FTOOLS software package. 
Source counts were extracted using a 5" radius for V, B and U filters
and a 12" for the UVW1, UVM2 and UVW2 filters. The background was
extracted from source-free regions in the surroundings. Filter UVW1
and UVM2 observations having four individual exposures, these were
stacked prior to aperture photometry as the {\tt uvotmaghist} task
showed no variability between the individual exposures. 
Count rates are then converted to fluxes using the standard
photometric zero points \cite{Poole2008}. The reported fluxes are
de-reddened for Galactic absorption following the description in
\cite{roming2009}, with $E(B-V) = 0.103$~mag.

\subsection{ATOM}
\begin{figure}[tb]
  \centering
  \includegraphics[width=0.7\textwidth]{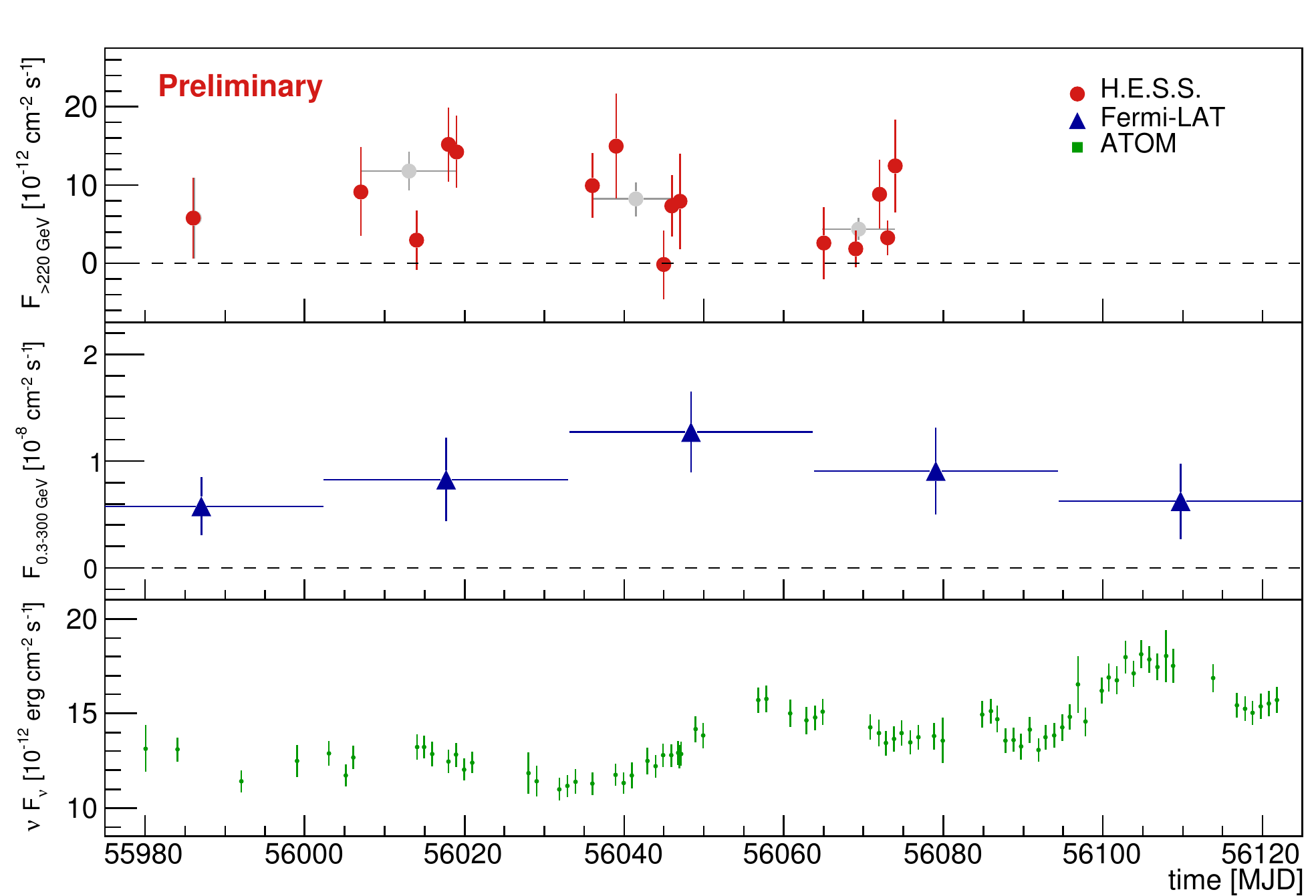}
  \caption{Light curve of \pks\ in different wavelength regimes. 
    {\it (top)} Nightly (red) and monthly (gray) binned fluxes above
    220~GeV measured by \hess\ 
    {\it (middle)} Monthly binned fluxes measured by \fermi-LAT. 
    {\it (bottom)} Nightly optical fluxes measured by ATOM in the R band. 
  }
  \label{fig:lc}
\end{figure}

The Automatic Telescope for Optical Monitoring (ATOM) is a fully
automatic 75cm optical telescope operated by LSW Heidelberg on the
\hess\ site \cite{hauser2004}. Operating since 2005, ATOM provides 
optical monitoring of gamma-ray sources and has observed \pks\ with
high cadence in 2012 in the R band. Following the VHE discovery,
additional B, V and J band observations were obtained. The ATOM 
data were analysed using the Automatic Data Reduction and Analysis
Software (ADRAS) which automates standard data
reduction and performs differential photometry using five known nearby
sources as calibrators. 
A preliminary light curve is presented in \reffig{fig:lc}, showing
clear variability in the optical regime contemporaneous with the
\hess\ observations.

\section{Spectral energy distribution and modeling} \label{sec:SED} 

A spectral energy distribution (SED) of \pks\ was derived using
the above described multi-wavelength data sets. As no significant
variability was detected in the gamma-ray regimes, the spectra
resulting from the analyses of the full data sets were used. 
Additionally, archival data from the 2MASS \cite{2mass} and WISE
catalogs \cite{wise} were used to complete the low-energy part of the SED. 

\begin{figure}[ht]
  \centering
  \includegraphics[width=0.75\textwidth]{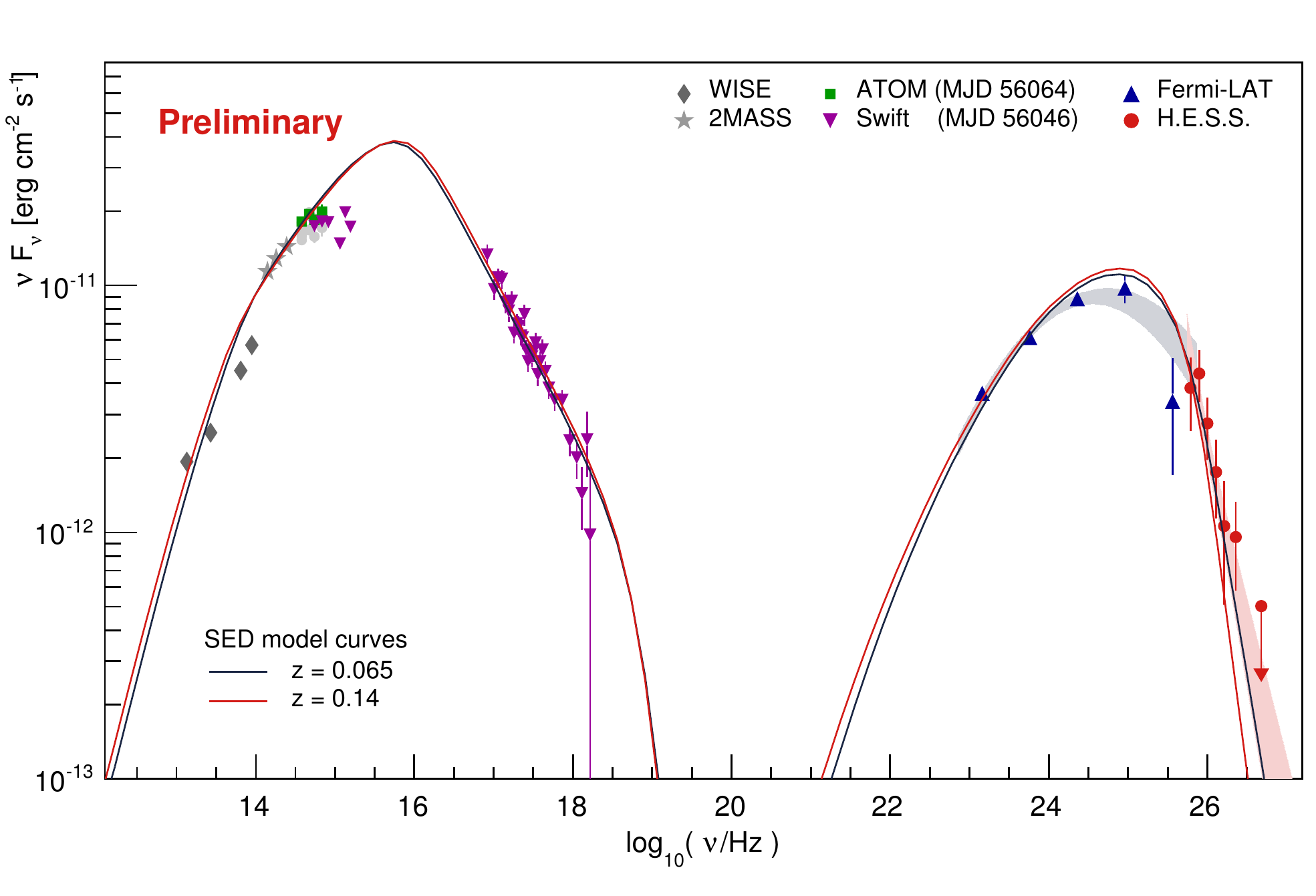}
  \caption{Spectral energy distribution of \pks\ together with the
    one-zone SSC model for two possible redshifts (see text for
    details). }
  \label{fig:SED}
\end{figure}

The overall, non-simultaneous SED of \pks\ is shown in
\reffig{fig:SED}. It exhibits the usual two bumps seen in 
blazars, one at low frequencies, from radio to X-rays, and the other
at higher frequencies. This SED was modeled with a one-zone leptonic
model \cite{boettcher2013}, taking the extragalactic absorption into
account \cite{finke2010}. Within the model the low-energy emission is 
interpreted as synchrotron emission from relativistic electrons in a
spherical emission region, moving relativistically along the jet
which is closely aligned with the line-of-sight. The high-energy
gamma-ray emission is produced via Compton up-scattering off the same
electron population which produced the synchrotron peak
(synchrotron-self Compton; SSC). 
The non-thermal electron distribution in the emission region is
determined self-consistently as an equilibrium between injection of a
power-law distribution, radiative cooling, and particle escape. 
Due to the uncertainty in the distance to \pks, the SED modeling was
done for two redshifts\footnote{Redshifts are converted to luminosity
  distances using a $\Lambda$CDM cosmology with
  $H_0=70\,\mathrm{km\,s}^{-1} \mathrm{Mpc}^{-1}$,
  $\Omega_{\mathrm{m}}=0.3$ and $\Omega_\Lambda=0.7$.}: 
$z=0.065$ and $z=0.14$. The results are shown in \reffig{fig:SED}. It
can be seen that the overall SED is moderately well described by the model for
the used parameter values which are given in \reftab{tab:SSC}. The
parameter values themselves are well within the range of those used for
previously detected VHE blazars \cite{heike} with a relatively low
magnetic field strength and a large emission region; resulting in a
particle dominated jet ($L_B/L_e < 0.01$, $L_B$ and $L_e$ being the
magnetic and kinetic jet power, respectively). Models with additional
external radiation fields yield, in general, jet energetics closer to
equipartition ($L_B/L_e \sim 1$), but due to the lack of additional
observational constrains, applying such a more complex model to the
SED of \pks\ would result in strongly underconstrained parameters and
seems therefore not justified.

\begin{table}[ht]
  \centering
  \begin{tabular}{lcccccccc cc}
  \hline \hline 
{$z$} & {$L_e$} & {$\gamma_1$} & {$\gamma_2$} & {$q$} & {$B $} & 
{$R $} & {$\Gamma$} & {$\eta_{\mathrm{esc}}$} & {$\epsilon_{Be}$} & 
{$\delta t_{\mathrm{var}}$} \\
{} & {[$10^{44}$~erg/s]} & {[$10^{3}$]} & {[$10^{5}$]} & {} & {[G]} & 
{[$10^{16}$~cm] } & {} & {} & {} & {[hr]} \\
\hline 
0.065 & 0.426 & 72 & 20 & 3.2 & 0.02 & 12 & 10 & 50 & 0.00507 & 86 \\
0.14  & 1.038 & 78 & 20 & 3.2 & 0.02 & 27 & 10 & 25 & 0.01054 & 207 \\
\hline 
\end{tabular}
\caption{SSC model parameters for \pks. 
The columns are the following: $z$ is the assumed redshift;
$L_{\mathrm{e}}$ is the (kinetic) jet luminosity; $\gamma_1$ is the low 
energy cutoff energy of the electron distribution; $\gamma_2$ is the
high energy cutoff; $q$ is the spectral index of the electron
injection spectrum; $B$ is the magnetic field strength; $R$ is the
emission region radius; $\Gamma$ is the bulk Lorentz factor; and
$\eta_{\mathrm{esc}}$ is the escape time parameter with
$t_{\mathrm{esc}} = \eta_{\mathrm{esc}} \cdot R/c$. Additionally, two
output parameters are given: $\epsilon_{Be}$ is the
resulting relative partition parameter $\epsilon_{Be} =
L_{\mathrm{B}}/L_{\mathrm{e}}$; and $\delta t_{\mathrm{var}}$ is the
resulting minimum variability time scale. }
  \label{tab:SSC} 
\end{table}

\section{Summary and Conclusions}

In this paper, we reported on the \hess\ discovery of the BL Lac
object \pks\ in 2012. The source was selected as VHE blazar candidate
based on its hard high-energy spectrum measured by \fermi-LAT, but is
also listed in \cite{2013ApJS..207...16M} based on its infrared and
X-ray properties. The average flux in the VHE regime was about 3\% of
the Crab Nebula flux and showed no significant gamma-ray variability. 
Multi-wavelength data, contemporaneous with the \hess\ observations,
were used to derive an SED which is well reproduced with a one-zone 
leptonic model with parameters typical for other VHE blazars. \\


{\small \noindent {\bf Acknowledgments:} 
{ 
The support of the Namibian authorities and of the University of Namibia in facilitating the construction and operation of H.E.S.S. is gratefully acknowledged, as is the support by the German Ministry for Education and Research (BMBF), the Max Planck Society, the German Research Foundation (DFG), the French Ministry for Research, the CNRS-IN2P3 and the Astroparticle Interdisciplinary Programme of the CNRS, the U.K. Science and Technology Facilities Council (STFC), the IPNP of the Charles University, the Czech Science Foundation, the Polish Ministry of Science and Higher Education, the South African Department of Science and Technology and National Research Foundation, and by the University of Namibia. We appreciate the excellent work of the technical support staff in Berlin, Durham, Hamburg, Heidelberg, Palaiseau, Paris, Saclay, and in Namibia in the construction and operation of the equipment.
}}

\end{document}